# Thermal boundary conductance of metal–diamond interfaces predicted by machine learning interatomic potentials

Khalid Zobaid Adnan[1], Mahesh R. Neupane[2], Tianli Feng[1*]

[1]Department of Mechanical Engineering, University of Utah, Salt Lake City, Utah 84112, United States

[2]Army Research Directorate (ARD), DEVCOM Army Research Laboratory, Adelphi, Maryland 20783, United States

Corresponding author: tianli.feng@utah.edu

**Abstract**:

Thermal boundary conductance (TBC) across metal–diamond interfaces plays a critical role in the thermal management of future diamond-based ultrawide bandgap semiconductor devices. Molecular dynamics is a sophisticated method to predict TBC but is limited by the lack of reliable potential describing metal–diamond interfaces. In this work, we report the development of machine learning interatomic potentials and the prediction of TBCs of several technologically promising metal–diamond interfaces using nonequilibrium molecular dynamics. The predicted TBCs of Al, Zr, Mo, and Au-diamond interfaces are approximately 316, 88, 52, and 55 MW/m$^2$K, respectively, after quantum corrections. The corresponding thermal boundary resistances are equivalent to 0.8-μm thick of Al, 1.4-μm Mo, 0.3-μm Zr, and 5.3-μm Au, respectively. We also find that the conventional simple models, such as the acoustic mismatch model and diffuse mismatch model, even including the full-band phonon dispersion from first principles, largely misestimate the TBC values because of their inability to include inelastic transmission as well as interfacial structural and bonding details. The quantum-corrected TBC values for the metal–diamond interfaces correlate well with the quantum-corrected phonon specific heat of metals, instead of diamond. Additionally, our comparative analysis of Debye temperature and elastic modulus in these systems reveals that the former parameter correlates more strongly with the TBC than the latter. These low TBC values need to be considered in future diamond-based semiconductor devices.

## 1. Introduction

Diamond is an ultrawide-bandgap (UWBG) material with[1] 5.5-eV bandgap, high carrier mobility[2,3], large breakdown field[4] (>10 MV/cm), low thermal expansion[5,6] and the highest thermal conductivity [7–10] (2200 W/m-K), which makes it an ideal material system for next-generation power and communication devices operating in extreme conditions[11–13]. High-electron-mobility transistors (HEMT)[14–16] result in significant self-heating[17,18] of devices. As the channel material in a transistor application, the diamond must be integrated with the metal electrodes[19]. In addition to the traditional downward heat dissipation through the diamond-substrate interface, an additional pathway upward through the metal–diamond interface could significantly contribute to the enhanced thermal management of diamond-based power electronics[20]. The thermal boundary resistance (TBR), which is the reciprocal of thermal boundary conductance (TBC), at the metal–diamond interfaces can impose substantial thermal resistance in these devices[21]. Gaining a comprehensive understanding of heat transfer at the clean metal–diamond interfaces is vital not only for ensuring the reliability and performance of diamond-based power devices, but also plays a key role in the broader context of thermal management[22–24]. Moreover, the metal–semiconductor interfaces are a common component not just in metal–oxide–semiconductor field-effect-transistor, but also in devices like junction field-effect transistors[25] and Schottky barrier diodes[26–28]. Hence, garnering a comprehensive understanding of these interfaces offers valuable insights and opportunities for the design and optimization of semiconductor devices.

TBCs of various metal–diamond interfaces have been measured in recent years. In Al–diamond interfaces, the reported time-domain thermoreflectance (TDTR) experimental TBC values are widely spread between 25 to 250 MW/m²K, attributing to the varying surface treatment conditions and surface orientations[29,30]. Mo–diamond interface also has a similar TBC range between 60–285 MW/m²K[31]. However, the interface includes a carbide formation ($MoC_2$) at the interface. In a subsequent study, Monachon and Weber[32] predicted the TBC for this interface to be around 220–240 with a partial carbide formation at the interface. In addition, the measured TBC of Au–diamond interface by Hohensee et al.[33] is ~134 MW/m²K, at a pressure of 34–35 GPa. It is worth noting that TBC at metal–diamond interfaces is influenced by pressure, as it has the potential to enhance interfacial bonding, thereby impacting the TBC across the interface. In a separate study by Blank et al.[34], interfaces with nanometer-thick interlayers, using either nickel or molybdenum, were investigated. The reported TBC, without any interlayer, was approximately 76 MW/m²K. The TBC of Au–diamond interfaces were reported to be about 40 MW/m²K by Stoner and Maris[35]. However, these experimental measurements were based on metal–diamond interfaces with various oxides, surface treatments, and carbide formations, altering the interfacial atomic structure. A fundamental understanding of the intrinsic TBCs in clean metal–diamond interfaces, a critical element in the diamond-based UWBG devices, is still limited.

Alongside experimental investigations, numerous theoretical studies have been undertaken to comprehend and predict the TBCs of metal–diamond interfaces. Lombard et al. reported a TBC of 130 MW/m²K for an Al–diamond interface using the diffuse mismatch model (DMM)[36]. In contrast, the reported values for TBC from DMM were 365 MW/m²K by Battabyal et al.[37]. Additionally, their work included the modified scattering-mediated acoustic mismatch model, which predicted a TBC of 225 MW/m²K. Similarly, the TBC of Au–diamond interfaces is predicted to be around 13 MW/m²K from DMM[36].

Though these theoretical studies provide some insights into the thermal properties, the theoretical models[38] used in these studies have their limitations. The models frequently oversimplify phonon transmission by neglecting inelastic contributions and may disregard inhomogeneous bonding configurations between dissimilar materials. Most of these models rely on the evaluation of phonon transmission coefficients at the interfaces. To calculate phonon transmission coefficients, common ways are the acoustic mismatch model (AMM)[39], DMM[40,41], and atomistic Green's function (AGF)[42–45]. The limitation of AMM and DMM is that they neglect the inelastic transmission and ignore the detailed atomic structure and bonding strength at the interface. The limitation of harmonic AGF is that it ignores inelastic transmission, and the limitation of the very recently developed anharmonic AGF is its extreme computational cost for complex interfaces with interlayer structures. Despite the improvement made by some works[46–54], they ignore some critical phenomena, such as the existence of interfacial phonon modes and the local phonon nonequilibrium[55]. The wave-packet method[56–60] is an effective way to study phonon transmission, but it usually studies one mode at a time, which is time-consuming. More importantly, it is typically performed at 0 K, assuming all the other modes are frozen out, which misses the phonon–phonon coupling effect at finite temperatures. Recently, mode-resolved phonon transmittance[61] using lattice dynamics with energy conservation, has been developed to study interfaces. Compared to the previously mentioned methods based on phonon transmission coefficients, molecular dynamics (MD) can naturally include all the physical phenomena[62] near the interfaces, along with the temperature effect[55]. Many spectral phonon analysis works used the MD method to gain insights of phonon transport[55,63–72]. However, the accuracy of MD solely relies on interatomic potentials[73], which are limited for most heterostructures. Though most of these potentials produce acceptable results for the intended models and applications, they are not transferrable and must be thoroughly benchmarked and validated before their use.

## 2. Materials and methods

In this work, we employ MD to predict the TBC across Al–, Zr–, Mo–, and Au–diamond interfaces, as shown in Fig. 1. But MD has three main limitations: (1) it severely depends on interatomic potentials, which are nonexistent for most heterostructures (and not accurate if they exist); (2) it uses classical Boltzmann instead of Bose–Einstein statistics; and (3) it cannot capture the electron–phonon coupling across the interface. To mitigate the first limitation, we develop accurate machine learning interatomic potentials (MLIPs) dedicated for the interface heterostructures based on ab initio molecular dynamics (AIMD) simulations. They are much more accurate than classical potentials, which were developed using bulk materials and mixing rules. To mitigate the second limitation of MD, we conduct a quantum correction by weighing the contributions of different phonon modes with the ratio of the quantum and classical heat capacities of that mode based on DMM transmission coefficients. Regarding the last limitation, electron–phonon[74] coupling is beyond the scope of this work. Nonetheless, electron–phonon coupling across interfaces has been found negligible for metal–dielectric interfaces[75–77].

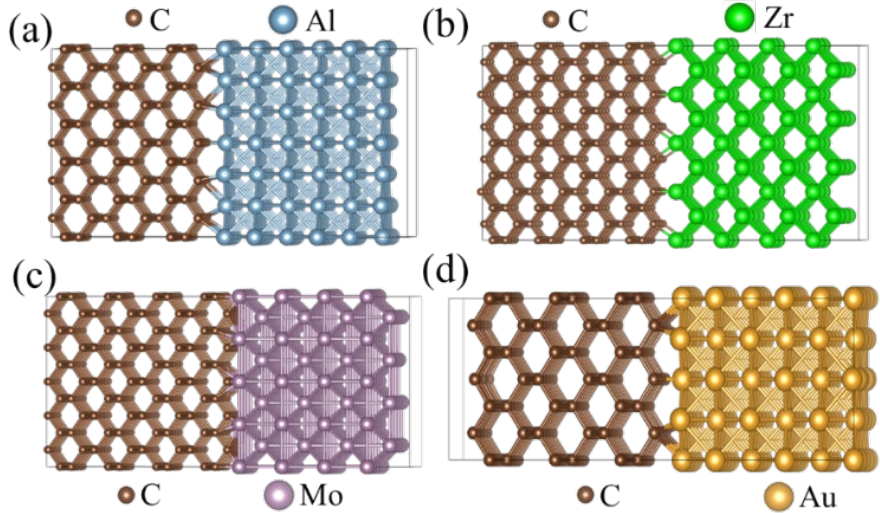

**FIG. 1**. Supercell atomic structures used in the AIMD simulations in this work for (a) Al–diamond, (b) Zr–diamond, (c) Mo–diamond, and (d) Au–diamond interfaces.

Because of the large in-plane lattice mismatch between the metal layers and underlying diamond substrate, construction of metal–diamond heterostructures is quite challenging. To overcome this and construct the most realistic heterostructure models, we created heterostructures by repeating metal layers and diamond substrate in X and Y directions. The heterostructures are chosen as 5×5 (Al) and 3×3 (diamond), 4×4 (Zr) and 2×2 (diamond), 4×4 (Mo) and 5×5 (diamond), and 4×4 (Au) and 3×3 (diamond), respectively, as shown in Fig. 1. The lattice mismatch will induce pressure laterally and induce strains on the metal side. The strain on diamond is negligible as it has a much larger modulus than metals. We first try to relax the lattice along the heat transport (axial) direction so there is no stress along the axial direction, but we find that this will induce phase change to the metals since the lateral directions have large stress. To preserve the metals' crystal symmetries, we let the strain along the axial direction be the same as the lateral directions. In the final structures, Al–, Zr–, Mo–, and Au–diamond heterostructures have isotropic pressures of 5–6-, 4.5–5-, 0.7–0.8-, and 34–35 GPa, respectively. We are aware of the expected overestimation of the TBCs of clean interfaces due to the intrinsic interface-induced pressure, and hence, we would like to reiterate that the results reported in this study will serve as an upper limit for the TBCs of these clean interfaces.

The density functional theory (DFT)-based AIMD simulations are conducted by using the Vienna Ab initio Simulation Package[78,79] (VASP) with the projector augmented wave (PAW)[80] pseudopotentials. The energies, stresses, atomic forces, and atomic configurations in the AIMD simulations are used to train the moment-tensor machine learning potentials (MTPs) by using the MLIP package[81]. The energy cutoffs are set to be 450 eV. The Γ point-only electronic **k**-mesh is used. For each interface, we run 2–6 independent AIMD simulations, each containing 1,000–1,500 steps at 300 K. Each simulation starts with random atomic positions and velocities. The timestep is 1 fs. The output files generated from AIMD contain energy, forces, and stresses for each timestep, which are used for MLIP training. The minimum and maximum cutoff radii of the MLIP are set at 1.0 and 6.0, 0.9 and 3.8, 0.96 and 5.0, and 0.92 and 3.8 Å for Al–, Zr–, Mo–, and Au–diamond interfaces, respectively. The training level[81] is set as 24. The number of iterations for training is set to 500. A total of 80%–90% of the data is allocated for the training, and the remaining portion is designated for the testing.

## 3. Results from machine learning interatomic potentials

The accuracy of the MLIPs is demonstrated by the root-mean-square error (RMSE) of the MLIP compared to DFT in Fig. 2. Generally, the forces predicted by the MLIP agree very well with those predicted directly from AIMD for all four structures.

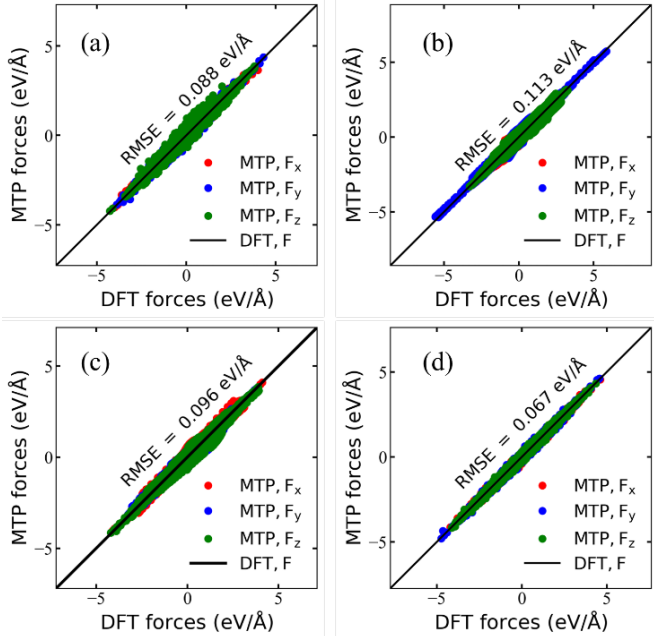

**FIG. 2:** Comparisons of forces on atoms between MTP and DFT calculations for (a) Al–diamond, (b) Zr–diamond, (c) Mo–diamond, and (d) Au–diamond interface systems.

With the trained MTPs, we run NEMD simulations using the Large-scale Atomic/Molecular Massively Parallel Simulator (LAMMPS)[82]. The atoms at the two edges are fixed to mimic the adiabatic boundary condition. The systems are first stabilized in constant volume and temperature (NVT) at 300 K for 3 million steps with a timestep of 1 fs. Then, the metal atoms next to the fixed boundary are changed to 320 K as the hot reservoir, and the diamond atoms next to the fixed boundary are changed to 280 K as the cold reservoir, both using the Langevin thermostat[83]. The number of atoms in both reservoirs is approximately 200. Then, the ensemble is changed to constant volume and energy (NVE) and runs for another 2 million steps. After that, the temperature and heat flux are recorded for another 2 million steps. The TBC is calculated by using $G = \frac{q}{A \cdot \Delta T}$, where $q$ is the heat flux, $\Delta T$ is the temperature jump at the interface, and $A$ is the cross-section area.

The TBCs predicted by machine learning molecular dynamics (MLMD) after quantum correction are shown in Fig. 3, compared to the available experimental data. The size effect of MD is studied but found to be not significant from 7.4 nm up to 51 nm. The predicted TBC values are 316 ± 22, 88 ± 8, 51 ± 8, and 55 ± 5 for Al–, Mo–, Zr–, and Au–diamond interfaces, respectively. The corresponding TBRs, assuming bulk thermal conductivity of metals, are equivalent to 0.75-μm thick of Al, 1.38-μm Mo, 0.30-μm Zr, and 5.28-μm Au, respectively. Though these values are negligible, for the wide bandgap semiconductor transistors with nano to micrometer scale they play major role in enhancing power-loss and thermal breakdown[84–87].

It is well established that the electron and phonon collectively contribute to the TBC across the interface. Quantitatively, the Al–diamond interface has the highest TBC value, followed by the Mo–diamond, Zr–diamond, and Au–diamond interfaces. The Al–diamond interface exhibits the least phonon mismatch compared to the other interfaces. This dominance of vibrational match contributes to its higher TBC compared to the other interfaces. The exception occurs between the Mo– and Zr–diamond interfaces. Despite the Zr–diamond exhibiting a smaller mass mismatch[88,89], it shows a lower TBC. This is because the Zr–diamond interface has a larger phonon mismatch than the Mo–diamond interface, which will be discussed later.

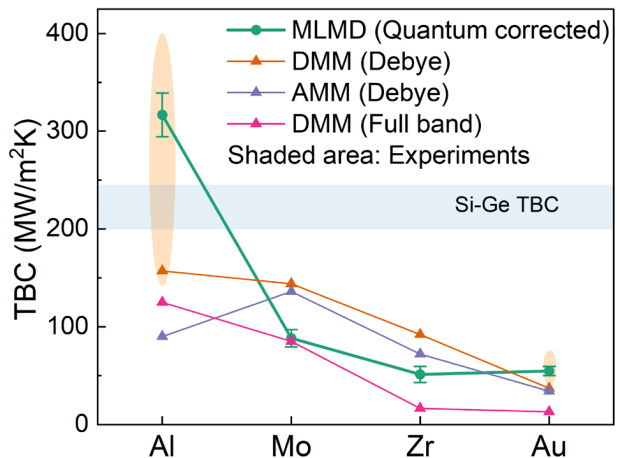

**FIG. 3:** TBC of Al–, Zr–, Mo–, and Au–diamond interfaces from MLMD (quantum corrected), AMM (Debye), DMM (Debye), and DMM (full band) models. Isotropic pressures of 5–6, 4.5–5, 0.7–0.8, and 34–35 GPa exist in the MLMD simulations, respectively. Yellow shaded area indicates the TBC measured for Al–diamond interface by Hohensee et al.[33] from 0 to 50 GPa. Blue shaded area represents the TBC of clean and interfacial mixing Si–Ge interface[90].

For the Al–diamond interface, the TBC values measured from the TDTR are widely spread across a range of 21–390 MW/m$^2$K[29,30,33,91]. It is mainly due to the surface quality and termination variability of the substrate before depositing Al. Monachen and Weber found that surface treatments and Al deposition techniques can also affect the TBC value[29,30], due to the change of cleanliness and surface termination of diamond. The hydrogen-treated diamond surfaces exhibit a 4× lower TBC value than the oxygenated diamond interfaces[91], mainly due to the differences in carrier scattering in the Al layer near the interfacial region. Hohensee et al.[33] measured that TBC can increase with pressure in a range of 150-400 MW/m$^2$K for 0-40 GPa. Our MLMD results reveal TBC of approximately 316 ± 22 MW/m$^2$K at an average pressure of about 5-6 GPa, which is slightly above the upper limit of the TBC data of 180-280 MW/m$^2$K measured by Hohensee et al.[33] at the same pressure range. The slightly higher MLMD result could be attributed to the ideal nature of the Al–diamond interface, without any impurities, doping, or vacancies. Interestingly, the Al–diamond TBC values are not much lower than that of clean Si–Ge interface, which is about 200–240 MW/m$^2$K from both experiment and simulations[90], despite that the Si–Ge interface has significantly better lattice and vibrational match, as well as chemical bonding, than the Al–diamond contact.

For the Mo–diamond interface, the predicted TBC of 88 ±8 MW/m²K is significantly smaller than the measured value of 220–240 MW/m$^2$K. This is because the measured data was based on Mo–$Mo_xC_y$ interface rather than a clean Mo–diamond interface[92]. The $Mo_xC_y$ (Mo/diamond composite) might contribute to a bridging effect[93,94], which can enhance TBC. Unfortunately, there is no experimental TBC data available for the clean Mo–diamond interface. Regarding the Zr–diamond interface, there is no experimental data available either.

For the Au–diamond interface, the experimental TBC from Hohensee et al.[33] is about 130 MW/m²K, observed at a pressure of about 30–35 GPa. It is much higher than the 55 ±5 MW/m²K obtained in MLMD simulations at a similar pressure of 34–36 GPa. This might be because of the following two reasons: (1) the measured data was for Au(Pd)–diamond interface, an alloy-like structure, or (2) the implemented surface treatment to enhance the binding between metal and diamond interfaces. In this study[33], without any pressure, the TBC was measured to be about 70 MW/m²K, which is very close to another reported TBC of 76 MW/m²K, measured by Blank et al.[34]. The insertion of a nanometer-thick Ni or Mo interlayer increases TBC to 195 ±40 MW/m²K[34] due to the bridging effect[93,94]. In another report by Stoner and Maris[35], the TBC of Au–diamond interface was measured to be about 40 MW/m²K, but detailed interfacial structure information is missing. In summary, the clean Au–diamond interface might be the reason for the smaller TBC value, which is smaller than 55 MW/m²K; however, adding interfacial mixing or interlayer in the implemented models might lead to the increased TBC value closer to the reported experimental values.

The interfaces between diamond and the other metals such as Cu, Pt, Pb, Nb, and W have also been extensively studied experimentally. These metal–diamond interfaces have a comparable TBC range as the four metal interfaces investigated in this study. For example, the TBC of Cu–diamond interface is about 33–73 MW/m²K[95] and can be enhanced to 96 MW/m²K with a Cr interlayer[96] and 87 MW/m²K with a Mo interlayer[97]. The TBC of Pt–diamond interface increases from 145 to 240 MW/m²K with pressure increasing from 0 to 50 GPa. The TBC of Pb–diamond interface increases from 30 to 140 MW/m²K from 0 to 50 GPa[33]. The TBC of Nb–diamond interface is about 70 MW/m²K, possibly with some carbide formation[98]. The TBC of W–diamond interfaces exhibit a wide range of 40 to 190 MW/m²K.

## 4. Results from DMM and AMM

The MLMD provides much more accurate predictions, compared to the conventional AMM- and DMM-based models. To make a comparison, we calculate the TBCs using DMM and AMM under the Debye approximation, as well as the DMM with the exact full-band phonon dispersion from first principles. Under the Debye model, the thermal conductance $G$ based on Landauer's formalism[99] is.

$$G_{\text{Debye}} = \frac{1}{4}\sum_j v_{A,j} \int_0^{\omega_{D,A}} \alpha_{A\to B} \hbar\omega \frac{df_{A,j}}{dT} d\omega, \qquad (1)$$

$$f_{A,j}(\omega, T) = \frac{\omega^2}{2\pi^2 v_{A,j}^3 [\exp\left(\frac{\hbar\omega}{k_B T}\right)-1]}, \qquad (2)$$

where $v_{A,j}$ is the sound velocity of material A in the polarization branch $j$, $\omega_{D,A}$ is the Debye frequency of material A calculated by $\omega_{D,A} = v_{A,j}(6\pi^2 n_A)^{\frac{1}{3}}$ with $n_A$ being the number density of atoms in material A, $\alpha_{A\to B}$ is the phonon transmission coefficient from material A to B, $\hbar$ is the reduced Planck constant, $\omega$ is the phonon angular frequency, $f_{A,j}$ is the phonon population under Debye approximation, and $k_B$ is the Boltzmann constant. $\alpha_{A\to B}$ can be obtained from AMM (assuming normal incident of phonons) and DMM using the following equations:

$$\alpha_{A\to B,\text{AMM}} = \frac{4 Z_A Z_B}{(Z_A + Z_B)^2}, \qquad (3)$$

$$\alpha_{A\to B,\text{DMM}} = \frac{\sum_j v_{B,j}^{-2}}{\sum_j v_{B,j}^{-2} + \sum_j v_{A,j}^{-2}}. \qquad (4)$$

Here, $Z_A$ and $Z_B$ are acoustic impedances of materials A and B, respectively. For DMM, we also conduct full-band Landauer's formalism [100–102] calculation by

$$G_{\text{DMM,full}} = \frac{1}{2(2\pi)^3}\sum_{j_A}\int_{BZ_A} c_{\lambda_A}\alpha_{A\to B}(\omega_{\lambda_A})|v_{\lambda_A,n}|d\mathbf{q}_A = \frac{1}{2N_{\mathbf{q},A}V_{c,A}}\sum_{j_A}\sum_{\mathbf{q}_A} c_{\lambda_A}\alpha_{A\to B}(\omega_{\lambda_A})|v_{\lambda_A,n}| \tag{5}$$

$$\alpha_{A\to B}(\omega) = \frac{\Delta\mathbf{q}_B \cdot \left(\sum_{\lambda_B}|v_{\lambda_B,n}|\cdot\delta_{\omega_{\lambda_B},\omega}\right)}{\Delta\mathbf{q}_A \cdot \left(\sum_{\lambda_A}|v_{\lambda_A,n}|\cdot\delta_{\omega_{\lambda_A},\omega}\right) + \Delta\mathbf{q}_B \cdot \left(\sum_{\lambda_B}|v_{\lambda_B,n}|\cdot\delta_{\omega_{\lambda_B},\omega}\right)}. \tag{6}$$

Here, $c_{\lambda_A} = k_B x^2 \frac{e^x}{(e^x-1)^2}$ is the specific heat of material A per mode, $x = \frac{\hbar\omega_{A,\lambda}}{k_B T}$, $v_{A,\lambda,n} = |\mathbf{v}_{A,\lambda}\cdot\mathbf{n}|$, $\mathbf{n}$ is the unit vector normal to the interface, $\mathbf{q}_A$ is the wavevector of material A, $\lambda_A \equiv (\mathbf{q}_A, j_A)$ is the phonon mode in material A with wavevector $\mathbf{q}_A$ and branch $j_A$, $\lambda_B \equiv (\mathbf{q}_B, j_B)$ is the phonon mode in material B, and $\Delta\mathbf{q}_A = \frac{V_{A,BZ}}{N_{\mathbf{q},A}}$ and $\Delta\mathbf{q}_B = \frac{V_{B,BZ}}{N_{\mathbf{q},B}}$ are the discretized Brillouin zone volume in material A and B, respectively. $V_{A,BZ}$ is the volume of the Brillouin zone of A, $N_{\mathbf{q},A}$ is the number of $\mathbf{q}$ points sampled in material A, and $V_{c,A}$ is the primitive cell volume of material A. Same for material B.

The results of AMM and DMM models' calculations are shown in Fig. 3. It is found that all the AMM, DMM, and DMM (full band) predictions are inaccurate and not reliable, compared to MLMD. DMM (full band) usually significantly underestimates the TBCs, because it neglects inelastic phonon transmissions[103]. Additionally, the AMM and DMM models do not account for the strength of atomic bonding or interfacial atomistic structure at the interface.

The limitation of MLMD is its classical nature of MD, which should be incorporated with quantum correction. To have a sense on the quantum correction of TBC, we plotted the $\text{TBC}_Q/\text{TBC}_C$ predicted from DMM (full band), where the subscript "Q" means the phonon specific heat is quantum in Eq. (5), i.e., $c = k_B x^2 \frac{e^x}{(e^x-1)^2}$, and the subscript "C" means the phonon specific heat is classical in Eq. (5), i.e., $c = k_B$. As shown in Fig 4(a), the ratio of $\text{TBC}_Q/\text{TBC}_C$ follows very well with the ratio of phonon specific heat $c_{ph,Q}/c_{ph,C}$ of the four metals. Obviously, $\text{TBC}_Q/\text{TBC}_C$ does not follow the ratio of phonon specific heat $c_{ph,Q}/c_{ph,C}$ of diamond, which has a high Debye temperature of 2200 K. This is expected, given that all the transmissions are elastic, and only the phonons below the low-cutoff frequency coming from the metal in the metal–diamond interface are included in the interfacial conductance. Therefore, at the first-order approximation, we use $c_Q/c_C$ as the quantum correction of the obtained TBC from MLMD, i.e., $G_{\text{MLMD},Q} = G_{\text{MLMD},C} \cdot \frac{c_{ph,Q,\text{metal}}(T)}{c_{ph,C,\text{metal}}}$. All the TBC values obtained from MLMD in this work are after quantum corrections. As shown in Fig. 4, the quantum correction of TBC at 300 K is very small since the phonon-specific heat of metals at 300 K is very close to the classical limit as the Debye temperatures of these metals are close to 300 K.

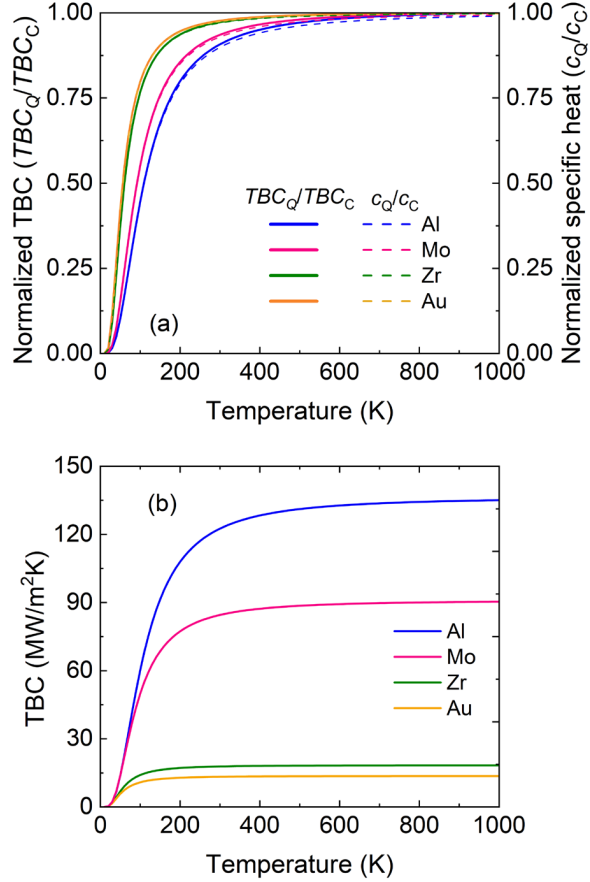

**FIG. 4:** Results from DMM (full band). (a) Effect of temperature on the TBC of Al–, Zr–, Mo–, Au–diamond interfaces. The classical TBC calculations imply the specific heat to be $k_B$, Boltzmann constant. The normalized specific heat ($c_Q/c_C$) is determined by taking the ratio of the metal's quantum specific heat and its classical specific heat.

## 5. Correlation between TBC and material properties

To find a correlation between TBC and the material properties in the predicted TBCs for the metal–diamond interfaces, we plot the TBC values as a function of atomic mass ratio, sound velocity ratio, Young's modulus ratio, and Debye temperature ratio, as shown in Fig. 5. Typically, the TBC should reach its maximum value when these ratios approach 1, which indicates that the two materials have similar properties resulting in continuous-like interface. In Fig. 5(a), we find that TBC increases with decreasing mass mismatch between metal and diamond, except for the Mo- and Zr-based interfaces. This is because, despite having a lighter mass and a more favorable mass correspondence with diamond compared to Mo, Zr exhibits a weaker bond and a greater disparity in bonding characteristics with diamond than Mo. Therefore, the TBC is not only determined by mass match but also bonding match. Therefore, the prediction of TBC values in terms of ascending or descending order cannot be qualitatively determined by mass or bonding mismatch. It is observed that Debye temperature mismatch between metal and diamond, serve well as a predictor of TBCs of metal–diamond interfaces. This is because Debye temperature ratio provides better understanding of bonding and mass mismatch, that is, phonon spectra mismatch, in the metal–diamond interfaces. Inspired by the results, we plot the available TBC values of various interfaces collected from the literature as functions of elastic modulus ratio and Debye temperature ratio, as shown in Fig. 6. The Debye temperature ratio is a better descriptor of the TBC than the elastic modulus ratio used in the literature[33,90,104]. We recommend using Debye temperature ratio, rather than elastic modulus ratio, as the indicator of TBC in the future.

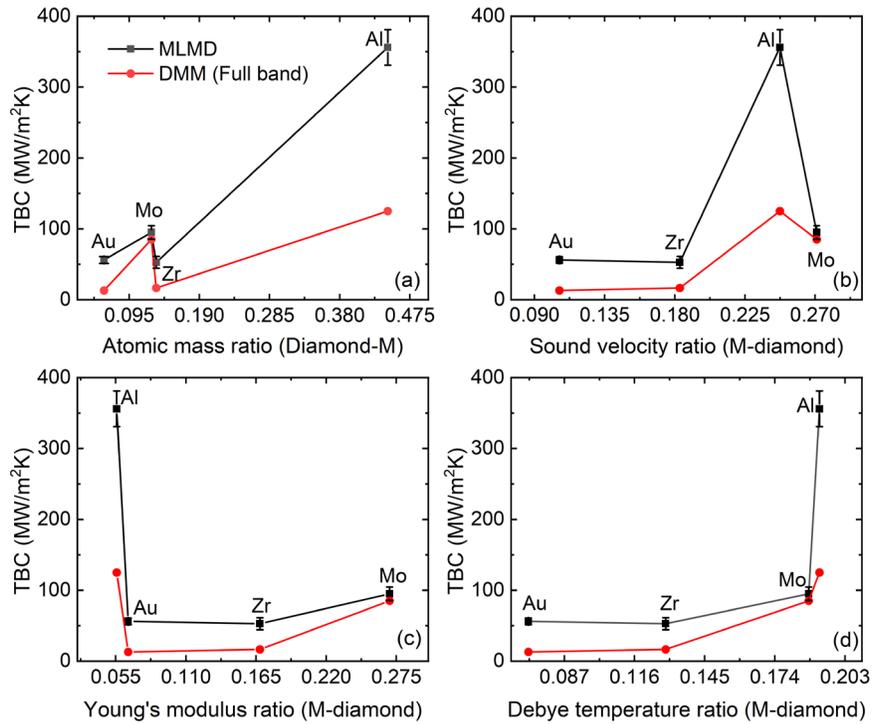

**FIG. 5:** TBCs at different metal–diamond interfaces as a function of (a) atomic mass ratio of diamond–M (b) sound velocity ratio of M–diamond (c) Young's modulus ratio of M–diamond and (d) Debye temperature ratio of M–diamond.

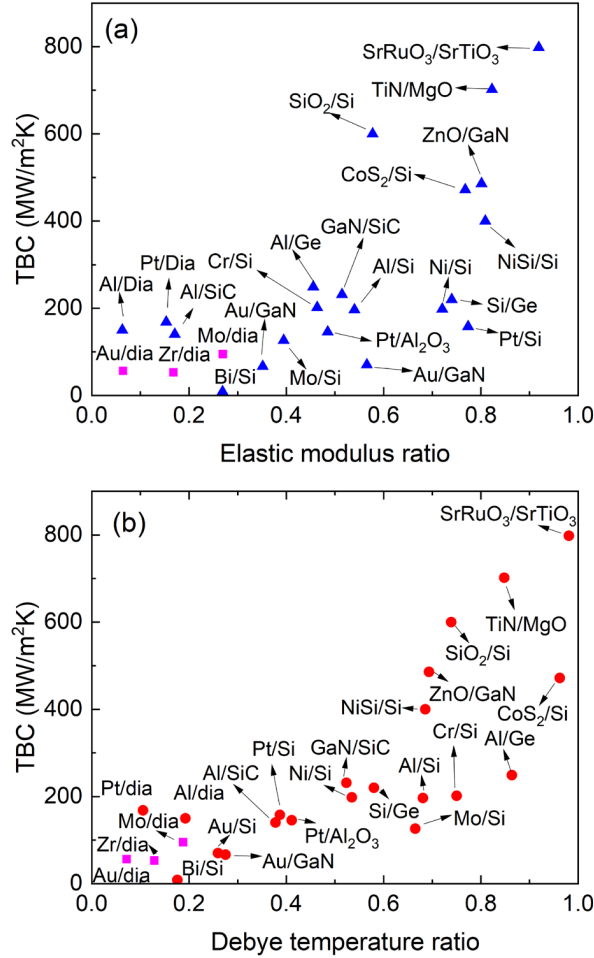

**FIG. 6:** TBCs for various interfaces collected from the literature[33,104] as a function of (a) elastic modulus ratio and (b) Debye temperature ratio. For clean Mo–, Zr–, and Au–diamond interfaces, which do not have available experimental data, the MLMD data are shown instead.

## 6. Conclusions

Diamond is being studied as the future ultimate wide-bandgap material worldwide, representing the frontier research in wide-bandgap technologies. The resistance between diamond and metal contacts (Al, Mo, Zr, and Au as most promising candidates), however, has not been simulated due to the lack of interatomic potential. This work fills the gap and provides timely guidance for the community. We have demonstrated the predictive capabilities of machine learning interatomic potential (MLIP) in predicting thermal boundary conductance (TBC) at Al–, Zr–, Mo–, and Au– diamond interfaces. Compared to the conventional AMM and DMM models, which only consider the elastic transmission, the MLMD simulations account for both elastic and inelastic transmissions, detailed interfacial bonding strength and structure, and all the other natural phonon transport mechanisms at the interface. The accuracy of MLIP is promising, as evidenced by low

RMSE values for forces between atoms. We observed that the Al–diamond interface has the largest, while the Zr–diamond interface poses lowest TBCs, indicating a high potential for employing Al metal as the electrode for diamond devices. These findings provide valuable fundamental insights into the interfacial thermal properties for future experimental studies in various metal–diamond interfaces.

**Declaration of Competing Interest**

The authors declare no conflicts of interest. The funders had no role in the design of the study; in the collection, analyses, or interpretation of data; in the writing of the manuscript; or in the decision to publish the results.

**Acknowledgment**

This work is supported by the National Science Foundation (NSF) (award number: CBET 2337749). We also acknowledge partial support from Oak Ridge Associated Universities (ORAU) Ralph E. Powe Junior Faculty Enhancement Awards. The support and resources from the Center for High Performance Computing at the University of Utah are gratefully acknowledged.

**Data Availability**

The data that support the findings of this study are available from the corresponding author upon reasonable request.

**References**

[1] M.H. Wong, O. Bierwagen, R.J. Kaplar, and H. Umezawa, "Ultrawide-bandgap semiconductors: An overview," Journal of Materials Research 2021 36:23 **36**(23), 4601–4615 (2021).

[2] J. Isberg, J. Hammersberg, E. Johansson, T. Wikström, D.J. Twitchen, A.J. Whitehead, S.E. Coe, and G.A. Scarsbrook, "High carrier mobility in single-crystal plasma-deposited diamond," Science **297**(5587), 1670–1672 (2002).

[3] J. Liu, H. Yu, S. Shao, J. Tu, X. Zhu, X. Yuan, J. Wei, L. Chen, H. Ye, and C. Li, "Carrier mobility enhancement on the H-terminated diamond surface," Diam Relat Mater **104**, 107750 (2020).

[4] H. Umezawa, "Recent advances in diamond power semiconductor devices," Mater Sci Semicond Process **78**, 147–156 (2018).

[5] P. Jacobson, and S. Stoupin, "Thermal expansion coefficient of diamond in a wide temperature range," Diam Relat Mater **97**, 107469 (2019).

[6] S. Stoupin, and Y. V. Shvyd'Ko, "Thermal expansion of diamond at low temperatures," Phys Rev Lett **104**(8), 085901 (2010).

[7] P.A. Loginov, D.A. Sidorenko, A.S. Orekhov, and E.A. Levashov, "A novel method for in situ TEM measurements of adhesion at the diamond–metal interface," Scientific Reports 2021 11:1 **11**(1), 1–10 (2021).

[8] G.T. Hohensee, R.B. Wilson, and D.G. Cahill, "Thermal conductance of metal–diamond interfaces at high pressure," Nat Commun **6**(1), 6578 (2015).

[9] B. Lee, J.S. Lee, S.U. Kim, K. Kim, O. Kwon, S. Lee, J.H. Kim, and D.S. Lim, "Simultaneous measurement of thermal conductivity and interface thermal conductance of diamond thin film," Journal of Vacuum Science & Technology B: Microelectronics and Nanometer Structures **27**(6), 2408 (2009).

[10] A. V. Inyushkin, "Thermal conductivity of group IV elemental semiconductors," J Appl Phys **134**(22), 221102 (2023).

[11] G. Perez, A. Maréchal, G. Chicot, P. Lefranc, P.O. Jeannin, D. Eon, and N. Rouger, "Diamond semiconductor performances in power electronics applications," Diam Relat Mater **110**, 108154 (2020).

[12] P.J. Wellmann, "Power Electronic Semiconductor Materials for Automotive and Energy Saving Applications – SiC, GaN, Ga2O3, and Diamond," Z Anorg Allg Chem **643**(21), 1312–1322 (2017).

[13] N. Donato, N. Rouger, J. Pernot, G. Longobardi, and F. Udrea, "Diamond power devices: state of the art, modelling, figures of merit and future perspective," J Phys D Appl Phys **53**(9), 093001 (2019).

[14] Y. Zhou, J. Anaya, J. Pomeroy, H. Sun, X. Gu, A. Xie, E. Beam, M. Becker, T.A. Grotjohn, C. Lee, and M. Kuball, "Barrier-layer optimization for enhanced GaN-on-diamond device cooling," ACS Appl Mater Interfaces **9**(39), 34416–34422 (2017).

[15] H. Guo, Y. Kong, and T. Chen, "Thermal simulation of high power GaN-on-diamond substrates for HEMT applications," Diam Relat Mater **73**, 260–266 (2017).

[16] Y.C. Hua, Y. Shen, Z.L. Tang, D.S. Tang, X. Ran, and B.Y. Cao, "Near-junction thermal managements of electronics," Adv Heat Transf **56**, 355–434 (2023).

[17] Y. Shen, and B.-Y. Cao, "Two-Temperature Principle for Electrothermal Performance Evaluation of GaN HEMTs," Appl Phys Lett **124**(4), (2023).

[18] Y. Shen, H.A. Yang, and B.Y. Cao, "Near-junction phonon thermal spreading in GaN HEMTs: A comparative study of simulation techniques by full-band phonon Monte Carlo method," Int J Heat Mass Transf **211**, 124284 (2023).

[19] S.P. Gimenez, "Diamond MOSFET: An innovative layout to improve performance of ICs," Solid State Electron **54**(12), 1690–1696 (2010).

[20] K. Yoshida, and H. Morigami, "Thermal properties of diamond/copper composite material," Microelectronics Reliability **44**(2), 303–308 (2004).

[21] T. Zhan, M. Xu, Z. Cao, C. Zheng, H. Kurita, F. Narita, Y.J. Wu, Y. Xu, H. Wang, M. Song, W. Wang, Y. Zhou, X. Liu, Y. Shi, Y. Jia, S. Guan, T. Hanajiri, T. Maekawa, A. Okino, and T. Watanabe, "Effects of Thermal Boundary Resistance on Thermal Management of Gallium-Nitride-Based Semiconductor Devices: A Review," Micromachines 2023, Vol. 14, Page 2076 **14**(11), 2076 (2023).

[22] N. Govindaraju, and R.N. Singh, "Processing of nanocrystalline diamond thin films for thermal management of wide-bandgap semiconductor power electronics," Materials Science and Engineering: B **176**(14), 1058–1072 (2011).

[23] D.S. Tang, and B.Y. Cao, "Phonon thermal transport and its tunability in GaN for near-junction thermal management of electronics: A review," Int J Heat Mass Transf **200**, 123497 (2023).

[24] Y. Xu, G. Wang, and Y. Zhou, "Broadly manipulating the interfacial thermal energy transport across the Si/4H-SiC interfaces via nanopatterns," Int J Heat Mass Transf **187**, 122499 (2022).

[25] K. Tsugawa, H. Umezawa, and H. Kawarada, "Characterization of diamond surface-channel metal-semiconductor field-effect transistor with device simulation," Japanese Journal of Applied Physics, Part 1: Regular Papers and Short Notes and Review Papers **40**(5 A), 3101–3107 (2001).

[26] M. Suzuki, S. Koizumi, M. Katagiri, T. Ono, N. Sakuma, H. Yoshida, T. Sakai, and S. Uchikoga, "Electrical characteristics of n-type diamond Schottky diodes and metal/diamond interfaces," Physica Status Solidi (a) **203**(12), 3128–3135 (2006).

[27] C. Cheng, Z. Zhang, X. Sun, Q. Gui, G. Wu, F. Dong, D. Zhang, Y. Guo, and S. Liu, "Ab-initio study of Schottky barrier heights at metal-diamond (1 1 1) interfaces," Appl Surf Sci **615**, 156329 (2023).

[28] P.K. Baumann, S.P. Bozeman, B.L. Ward, and R.J. Nemanich, "Characterization of metal-diamond interfaces: Electron affinity and Schottky barrier height," Diam Relat Mater **6**(2–4), 398–402 (1997).

[29] C. Monachon, and L. Weber, "Effect of diamond surface orientation on the thermal boundary conductance between diamond and aluminum," Diam Relat Mater **39**, 8–13 (2013).

[30] C. Monachon, and L. Weber, "Influence of diamond surface termination on thermal boundary conductance between Al and diamond," J Appl Phys **113**(18), 183504 (2013).

[31] C. Monachon, and L. Weber, "Thermal boundary conductance of transition metals on diamond," Emerging Materials Research **1**(2), 89–98 (2012).

[32] C. Monachon, and L. Weber, "Thermal boundary conductance between refractory metal carbides and diamond," Acta Mater **73**, 337–346 (2014).

[33] G.T. Hohensee, R.B. Wilson, and D.G. Cahill, "Thermal conductance of metal–diamond interfaces at high pressure," Nature Communications 2015 6:1 **6**(1), 1–9 (2015).

[34] M. Blank, G. Schneider, J. Ordonez-Miranda, and L. Weber, "Role of the electron-phonon coupling on the thermal boundary conductance of metal/diamond interfaces with nanometric interlayers," J Appl Phys **126**(16), (2019).

[35] R.J. Stoner, and H.J. Maris, "Kapitza conductance and heat flow between solids at temperatures from 50 to 300 K," Phys Rev B **48**(22), 16373 (1993).

[36] J. Lombard, F. Detcheverry, and S. Merabia, "Influence of the electron–phonon interfacial conductance on the thermal transport at metal/dielectric interfaces," Journal of Physics: Condensed Matter **27**(1), 015007 (2014).

[37] M. Battabyal, O. Beffort, S. Kleiner, S. Vaucher, and L. Rohr, "Heat transport across the metal–diamond interface," Diam Relat Mater **17**(7–10), 1438–1442 (2008).

[38] H. Bao, J. Chen, X. Gu, and B. Cao, "A Review of Simulation Methods in Micro/Nanoscale Heat Conduction," ES Energy & Environment, (2018).

[39] W.A. Little, "THE TRANSPORT OF HEAT BETWEEN DISSIMILAR SOLIDS AT LOW TEMPERATURES," Can J Phys **37**(3), 334–349 (1959).

[40] E.T. Swartz, and R.O. Pohl, "Thermal resistance at interfaces," Appl Phys Lett **51**(26), 2200–2202 (1987).

[41] E. Swartz, and R. Pohl, "Thermal boundary resistance," Rev Mod Phys **61**(3), 605–668 (1989).

[42] W. Zhang, N. Mingo, and T. Fisher, "Simulation of phonon transport across a non-polar nanowire junction using an atomistic Green's function method," Phys Rev B **76**(19), 195429 (2007).

[43] W. Zhang, T.S. Fisher, and N. Mingo, "Simulation of Interfacial Phonon Transport in Si–Ge Heterostructures Using an Atomistic Green's Function Method," J Heat Transfer **129**(4), 483 (2007).

[44] N. Mingo, and L. Yang, "Phonon transport in nanowires coated with an amorphous material: An atomistic Green's function approach," Phys Rev B **68**(24), 245406 (2003).

[45] Z. Tian, K. Esfarjani, and G. Chen, "Enhancing phonon transmission across a Si/Ge interface by atomic roughness: First-principles study with the Green's function method," Phys Rev B **86**(23), 235304 (2012).

[46] P.E. Hopkins, and P.M. Norris, "Effects of joint vibrational states on thermal boundary conductance," Nanoscale and Microscale Thermophysical Engineering **11**(3–4), 247–257 (2007).

[47] P.E. Hopkins, and P.M. Norris, "Relative contributions of inelastic and elastic diffuse phonon scattering to thermal boundary conductance across solid interfaces," J Heat Transfer **131**(2), 022402 (2009).

[48] P.E. Hopkins, J.C. Duda, and P.M. Norris, "Anharmonic phonon interactions at interfaces and contributions to thermal boundary conductance," J Heat Transfer **133**(6), 062401 (2011).

[49] R. Prasher, X. Hu, Y. Chalopin, N. Mingo, K. Lofgreen, S. Volz, F. Cleri, and P. Keblinski, "Turning Carbon Nanotubes from Exceptional Heat Conductors into Insulators," Phys Rev Lett **102**(10), 105901 (2009).

[50] P. Reddy, K. Castelino, and A. Majumdar, "Diffuse mismatch model of thermal boundary conductance using exact phonon dispersion," Appl Phys Lett **87**(21), 1–3 (2005).

[51] S. Sadasivam, N. Ye, J.P. Feser, J. Charles, K. Miao, T. Kubis, and T.S. Fisher, "Thermal transport across metal silicide-silicon interfaces: First-principles calculations and Green's function transport simulations," Phys Rev B **95**(8), 085310 (2017).

[52] S. Volz, editor, *Thermal Nanosystems and Nanomaterials* (Springer Berlin Heidelberg, Berlin, Heidelberg, 2009).

[53] S. Shin, M. Kaviany, T. Desai, and R. Bonner, "Roles of atomic restructuring in interfacial phonon transport," Phys Rev B **82**(8), 081302 (2010).

[54] J. Dai, and Z. Tian, "Rigorous formalism of anharmonic atomistic Green's function for three-dimensional interfaces," Phys Rev B **101**(4), 41301 (2020).

[55] T. Feng, Y. Zhong, J. Shi, and X. Ruan, "Unexpected high inelastic phonon transport across solid-solid interface: Modal nonequilibrium molecular dynamics simulations and Landauer analysis," Phys Rev B **99**(4), 045301 (2019).

[56] P.K. Schelling, S.R. Phillpot, and P. Keblinski, "Kapitza conductance and phonon scattering at grain boundaries by simulation," J Appl Phys **95**(11 I), 6082–6091 (2004).

[57] J. Shi, J. Lee, Y. Dong, A. Roy, T.S. Fisher, and X. Ruan, "Dominant phonon polarization conversion across dimensionally mismatched interfaces: Carbon-nanotube–graphene junction," Phys Rev B **97**(13), 134309 (2018).

[58] N.A. Roberts, and D.G. Walker, "Phonon wave-packet simulations of Ar/Kr interfaces for thermal rectification," J Appl Phys **108**(12), 123515 (2010).

[59] C.H. Baker, D.A. Jordan, and P.M. Norris, "Application of the wavelet transform to nanoscale thermal transport," Phys Rev B **86**(10), 104306 (2012).

[60] P.K. Schelling, S.R. Phillpot, and P. Keblinski, "Phonon wave-packet dynamics at semiconductor interfaces by molecular-dynamics simulation," Appl Phys Lett **80**(14), 2484 (2002).

[61] H.A. Yang, and B.Y. Cao, "Mode-resolved phonon transmittance using lattice dynamics: Robust algorithm and statistical characteristics," J Appl Phys **134**(15), 155302 (2023).

[62] Y. Xu, L. Yang, and Y. Zhou, "The interfacial thermal conductance spectrum in nonequilibrium molecular dynamics simulations considering anharmonicity, asymmetry and quantum effects," Physical Chemistry Chemical Physics **24**(39), 24503–24513 (2022).

[63] Y. Chalopin, and S. Volz, "A microscopic formulation of the phonon transmission at the nanoscale," Appl Phys Lett **103**(5), 051602 (2013).

[64] K. Sääskilahti, J. Oksanen, J. Tulkki, and S. Volz, "Role of inelastic phonon scattering in the spectrally decomposed thermal conductance at interfaces," Phys Rev B **90**(13), 134312 (2014).

[65] K. Sääskilahti, J. Oksanen, J. Tulkki, and S. Volz, "Spectral mapping of heat transfer mechanisms at liquid-solid interfaces," Phys Rev E **93**(5), 052141 (2016).

[66] Y. Zhou, and M. Hu, "Full quantification of frequency-dependent interfacial thermal conductance contributed by two- and three-phonon scattering processes from nonequilibrium molecular dynamics simulations," Phys Rev B Condens Matter Mater Phys **95**(11), 115313 (2017).

[67] T. Murakami, T. Hori, T. Shiga, and J. Shiomi, "Probing and tuning inelastic phonon conductance across finite-thickness interface," Applied Physics Express **7**(12), 121801 (2014).

[68] A. Giri, J.L. Braun, and P.E. Hopkins, "Implications of interfacial bond strength on the spectral contributions to thermal boundary conductance across solid, liquid, and gas interfaces: A molecular dynamics study," Journal of Physical Chemistry C **120**(43), 24847–24856 (2016).

[69] K. Gordiz, and A. Henry, "A formalism for calculating the modal contributions to thermal interface conductance," New J Phys **17**(10), 103002 (2015).

[70] K. Gordiz, and A. Henry, "Phonon transport at interfaces: Determining the correct modes of vibration," J Appl Phys **119**(1), 015101 (2016).

[71] K. Gordiz, and A. Henry, "Phonon Transport at Crystalline Si/Ge Interfaces: The Role of Interfacial Modes of Vibration," Sci Rep **6**, 23139 (2016).

[72] Y.X. Xu, H.Z. Fan, and Y.G. Zhou, "Quantifying spectral thermal transport properties in framework of molecular dynamics simulations: a comprehensive review," Rare Metals 2023 42:12 **42**(12), 3914–3944 (2023).

[73] C. Si, X.D. Wang, Z. Fan, Z.H. Feng, and B.Y. Cao, "Impacts of potential models on calculating the thermal conductivity of graphene using non-equilibrium molecular dynamics simulations," Int J Heat Mass Transf **107**, 450–460 (2017).

[74] Y. Xu, H. Fan, Z. Li, and Y. Zhou, "Signatures of anharmonic phonon transport in ultrahigh thermal conductance across atomically sharp metal/semiconductor interface," Int J Heat Mass Transf **201**, 123628 (2023).

[75] Z. Cheng, Y.R. Koh, H. Ahmad, R. Hu, J. Shi, M.E. Liao, Y. Wang, T. Bai, R. Li, E. Lee, E.A. Clinton, C.M. Matthews, Z. Engel, L. Yates, T. Luo, M.S. Goorsky, W.A. Doolittle, Z. Tian, P.E. Hopkins, and S. Graham, "Thermal conductance across harmonic-matched epitaxial Al-sapphire heterointerfaces," Communications Physics 2020 3:1 **3**(1), 1–8 (2020).

[76] R.E. Jones, J.C. Duda, X.W. Zhou, C.J. Kimmer, and P.E. Hopkins, "Investigation of size and electronic effects on Kapitza conductance with non-equilibrium molecular dynamics," Appl Phys Lett **102**(18), 183119 (2013).

[77] Y.R. Koh, M.S. Bin Hoque, H. Ahmad, D.H. Olson, Z. Liu, J. Shi, Y. Wang, K. Huynh, E.R. Hoglund, K. Aryana, J.M. Howe, M.S. Goorsky, S. Graham, T. Luo, J.K. Hite, W.A. Doolittle, and P.E. Hopkins, "High thermal conductivity and thermal boundary conductance of homoepitaxially grown gallium nitride (GaN) thin films," Phys Rev Mater **5**(10), 104604 (2021).

[78] G. Kresse, and J. Furthmüller, "Efficiency of ab-initio total energy calculations for metals and semiconductors using a plane-wave basis set," Comput Mater Sci **6**(1), 15–50 (1996).

[79] G. Kresse, and D. Joubert, "From ultrasoft pseudopotentials to the projector augmented-wave method," Phys Rev B **59**(3), 1758 (1999).

[80] P.E. Blöchl, "Projector augmented-wave method," Phys Rev B **50**(24), 17953 (1994).

[81] M.J. Waters, J.M. Rondinelli, I.S. Novikov, K. Gubaev, E. V Podryabinkin, and A. V Shapeev, "The MLIP package: moment tensor potentials with MPI and active learning," Mach Learn Sci Technol **2**(2), 025002 (2020).

[82] S. Plimpton, "Fast Parallel Algorithms for Short-Range Molecular Dynamics," Other Information: PBD: May 1993, (1993).

[83] Y. Hu, T. Feng, X. Gu, Z. Fan, X. Wang, M. Lundstrom, S.S. Shrestha, and H. Bao, "Unification of nonequilibrium molecular dynamics and the mode-resolved phonon Boltzmann equation for thermal transport simulations," Phys Rev B **101**(15), 155308 (2020).

[84] M.J. Fang, C.W. Tsao, and Y.J. Hsu, "Semiconductor nanoheterostructures for photoconversion applications," J Phys D Appl Phys **53**(14), 143001 (2020).

[85] B. Radisavljevic, A. Radenovic, J. Brivio, V. Giacometti, and A. Kis, "Single-layer MoS2 transistors," Nature Nanotechnology 2011 6:3 **6**(3), 147–150 (2011).

[86] C. Qiu, Z. Zhang, M. Xiao, Y. Yang, D. Zhong, and L.M. Peng, "Scaling carbon nanotube complementary transistors to 5-nm gate lengths," Science (1979) **355**(6322), 271–276 (2017).

[87] A.D. Franklin, and Z. Chen, "Length scaling of carbon nanotube transistors," Nature Nanotechnology 2010 5:12 **5**(12), 858–862 (2010).

[88] B. Xu, S. Hu, S.W. Hung, C. Shao, H. Chandra, F.R. Chen, T. Kodama, and J. Shiomi, "Weaker bonding can give larger thermal conductance at highly mismatched interfaces," Sci Adv **7**(17), 8197–8220 (2021).

[89] M. Masuduzzaman, and B. Kim, "Scale Effects in Nanoscale Heat Transfer for Fourier's Law in a Dissimilar Molecular Interface," ACS Omega **5**(41), 26527–26536 (2020).

[90] Z. Cheng, R. Li, X. Yan, G. Jernigan, J. Shi, M.E. Liao, N.J. Hines, C.A. Gadre, J.C. Idrobo, E. Lee, K.D. Hobart, M.S. Goorsky, X. Pan, T. Luo, and S. Graham, "Experimental observation of localized interfacial phonon modes," Nature Communications 2021 12:1 **12**(1), 1–10 (2021).

[91] K.C. Collins, S. Chen, and G. Chen, "Effects of surface chemistry on thermal conductance at aluminum-diamond interfaces," Appl Phys Lett **97**(8), 83102 (2010).

[92] C. Monachon, and L. Weber, "Thermal boundary conductance of transition metals on diamond," Https://Doi.Org/10.1680/Emr.11.00011 **1**(2), 89–98 (2015).

[93] T. Feng, Y. Zhong, J. Shi, and X. Ruan, "Unexpected high inelastic phonon transport across solid-solid interface: Modal nonequilibrium molecular dynamics simulations and Landauer analysis," Phys Rev B **99**(4), 045301 (2019).

[94] R. Xie, J. Tiwari, and T. Feng, "Impacts of various interfacial nanostructures on spectral phonon thermal boundary conductance," J Appl Phys **132**(11), 115108 (2022).

[95] V. Sinha, J.J. Gengler, C. Muratore, and J.E. Spowart, "Effects of disorder state and interfacial layer on thermal transport in copper/diamond system," J Appl Phys **117**(7), 74305 (2015).

[96] X. Liu, F. Sun, L. Wang, Z. Wu, X. Wang, J. Wang, M.J. Kim, and H. Zhang, "The role of Cr interlayer in determining interfacial thermal conductance between Cu and diamond," Appl Surf Sci **515**, 146046 (2020).

[97] G. Chang, F. Sun, L. Wang, Y. Zhang, X. Wang, J. Wang, M.J. Kim, and H. Zhang, "Mo-interlayer-mediated thermal conductance at Cu/diamond interface measured by time-domain thermoreflectance," Compos Part A Appl Sci Manuf **135**, 105921 (2020).

[98] C. Monachon, and L. Weber, "Thermal boundary conductance of transition metals on diamond," Emerging Materials Research **1**(2), 89–98 (2012).

[99] E.T. Swartz, and R.O. Pohl, "Thermal boundary resistance," (n.d.).

[100] P. Reddy, K. Castelino, and A. Majumdar, "Diffuse mismatch model of thermal boundary conductance using exact phonon dispersion," Appl Phys Lett **87**(21), 1–3 (2005).

[101] H. Subramanyan, K. Kim, T. Lu, J. Zhou, and J. Liu, "On the importance of using exact full phonon dispersions for predicting interfacial thermal conductance of layered materials using diffuse mismatch model," AIP Adv **9**(11), 115116 (2019).

[102] J. Zhou, B. Liao, G. Chen, J. Maassen, and M. Lundstrom, "(Invited) The Landauer Approach to Electron and Phonon Transport," ECS Trans **69**(9), 23 (2015).

[103] T. Feng, Y. Zhong, J. Shi, and X. Ruan, "Unexpected high inelastic phonon transport across solid-solid interface: Modal nonequilibrium molecular dynamics simulations and Landauer analysis," Phys Rev B **99**(4), 045301 (2019).

[104] A. Giri, P.E. Hopkins, A. Giri, and E. Hopkins, "A Review of Experimental and Computational Advances in Thermal Boundary Conductance and Nanoscale Thermal Transport across Solid Interfaces," Adv Funct Mater **30**(8), 1903857 (2020).